\documentclass[%
 aip,
 amsmath,amssymb,floatfix,
preprint,%
]{revtex4-1}

\usepackage{graphicx}
\usepackage{dcolumn}
\usepackage{bm}
\usepackage[mathlines]{lineno}

\usepackage[version=4]{mhchem}
\usepackage{cleveref,xcolor}
\usepackage{siunitx,mleftright}
\mleftright 

\usepackage[utf8]{inputenc}
\usepackage[T1]{fontenc}
\usepackage{mathptmx}

\DeclareSIUnit[]\kT
{\ensuremath{\mathnormal{k_\text{B}T}}}

\newcommand{\lab}[1]{{\bf{#1}}}
\newcommand{\stw}{0.65}

\newcommand{\comment}[1]{{#1}}

\begin{document}

\preprint{AIP/123-QED}

\title{Repulsive and attractive depletion forces mediated by nonadsorbing polyelectrolytes in the Donnan limit}

\author{Jasper Landman}
\affiliation{Laboratory of Physical Chemistry, Department of Chemical Engineering and Chemistry, Eindhoven University of Technology, P.O. Box 513, 5600 MB Eindhoven, the Netherlands}
\affiliation{Laboratory of Physics and Physical Chemistry of Foods, Wageningen University \& Research, P.O. Box 17, Bornse Weilanden 9, 6708 WG Wageningen, the Netherlands}
\author{Max P. M. Schelling}
\affiliation{Laboratory of Physical Chemistry, Department of Chemical Engineering and Chemistry, Eindhoven University of Technology, P.O. Box 513, 5600 MB Eindhoven, the Netherlands}
\author{Remco Tuinier}
\email{r.tuinier@tue.nl}
\affiliation{Laboratory of Physical Chemistry, Department of Chemical Engineering and Chemistry, Eindhoven University of Technology, P.O. Box 513, 5600 MB Eindhoven, the Netherlands}
\affiliation{Institute for Complex Molecular Systems, Eindhoven University of Technology, P.O. Box 513, 5600 MB Eindhoven, the Netherlands}

\author{Mark Vis}
\email{m.vis@tue.nl}
\affiliation{Laboratory of Physical Chemistry, Department of Chemical Engineering and Chemistry, Eindhoven University of Technology, P.O. Box 513, 5600 MB Eindhoven, the Netherlands}
\affiliation{Institute for Complex Molecular Systems, Eindhoven University of Technology, P.O. Box 513, 5600 MB Eindhoven, the Netherlands}

\date{\today}

\begin{abstract}
In mixtures of colloids and nonadsorbing polyelectrolytes, a Donnan potential arises across the region between surfaces that are depleted of polyelectrolyte and the rest of the system. This Donnan potential tends to shift the polyelectrolyte density profile towards the colloidal surface and leads to local accumulation of polyelectrolytes. We derive a zero-field theory for the disjoining pressure between two parallel flat plates. Polyelectrolyte is allowed to enter the confined interplate region at the cost of a conformational free energy penalty. The resulting disjoining pressure shows a crossover to a repulsive regime when the interplate separation gets smaller than the size of the polyelectrolyte chain, followed by an attractive part. We find a quantitative match between the model and self-consistent field computations that take into account the full Poisson-Boltzmann electrostatics.
\end{abstract}

\maketitle

\section{Introduction}
In both biological systems and industrial products, colloids or nanoparticles are often mixed with polymers in a common solvent. In biochemistry the focus has traditionally been on studying the properties of biologically relevant compounds like enzymes, in dilute solution at isotonic conditions with respect to the ionic strength; the presence of particles and macromolecules has only came into focus more recently: there is increasing attention for the effects of what is called ‘macromolecular crowding’ \cite{goodsell1991inside,ellis2001macromolecular,zimmerman1991estimation,minton2006can} on for instance enzyme activity \cite{norris2011true}. This led to the insight that perhaps even the organization of compartments within cells is driven by macromolecular crowding \cite{hancock2004role,Brangwynne2013PhaseOrganelles}.

In industrial products such as paints \cite{deWith}, cosmetics \cite{tadros2008colloids} or food products \cite{doublier2000protein}, polymers are also often mixed with colloids; the phase stability of these mixtures is a key issue that needs to be under control. In many cases there are no attractive interactions between the polymeric compounds and the colloidal particles in colloid--polymer mixtures. In such cases excluded volume interactions dominate the phase stability. 
Also mixtures featuring charged colloids and/or charged polymers offer the opportunity to generate complex crystal structures, including colloidal alloy and eutectic structures. 
For instance Toyotama \textit{et al}.\cite{toyotama2016spontaneous} reported the formation of area-filling eutectic structures in dispersions containing  charged colloidal particles and polyelectrolyte chains. 

To quantify the interactions and resulting phase behaviour in colloid--polymer mixtures, the concept of depletion-induced attraction has been used frequently and has been quite successful to describe well-defined systems \cite{fleer2008analytical}. Asakura and Oosawa \cite{asakura1954interaction,asakura1958interaction} were the first to realize that nonadsorbing polymer chains induce an indirect attraction between colloidal particles. Near the nonadsorbing surfaces a so-called depletion zone exists, where the polymer concentration is lower than in the bulk of the solution. The extent of this region is approximately the radius of gyration of the polymer in the dilute regime, whereas it equals the correlation length in the semi-dilute regime. Once depletion zones around two colloids overlap, an osmotic force becomes operational which pushes the colloidal particles together. This effective attraction, mediated by excluded volume repulsions, can lead to phase transitions, aggregation, or gelation when exceeding a certain polymer concentration. The focus of most work related to the depletion-mediated attraction has been on well-defined neutral systems \cite{poon2002physics}. 

In practice, however, either the polymer chains and/or the colloidal particles often carry charges, especially when dispersed in polar solvents such as water. \comment{For instance, in food products phase stability is of key relevance. Food dispersions often contain protein (aggregates), oil droplets and/or polysaccharides \cite{GRINBERG1997145,doublier2000protein,SYRBE1998179}. In many cases these food colloids contain (often negatively) charged polysaccharides and proteins. Similarly, water-based paint formulations contain a wide variety of particles and polymers, such as pigments and binders, which may be charged \cite{deWith,Tadros:2011wl}. Hence it is useful to develop more understanding of the role of polyelectrolytes on the phase stability of such colloid--polyelectrolyte mixtures.} Still, the understanding of the effects of charges in mixtures of colloids and polymers is rather limited from a theoretical point of view, even though Asakura and Oosawa discussed the possible effects of charges already at an early stage \cite{asakura1958interaction}. Below we briefly mention some earlier findings. 

There is quite a number of studies on charged polymers mixed with charged colloids. Already in the 1970s Kose and Hachisu \cite{kose1976ordered} studied the effect of the addition of sodium polyacrylate to polystyrene latex particles (both negatively charged) and found that this leads to crystallization of the colloidal spheres. This must originate from enhanced excluded volume interactions mediated by the polyelectrolytes. 
B\"{o}hmer \textit{et al}. \cite{bohmer1990weak} applied numerical lattice computations based upon the self-consistent field method of Scheutjens and Fleer \cite{scheutjens1979statistical,scheutjens1982effect,fleer1993polymers} to study the polyelectrolyte-mediated effect on the interaction between flat walls.
Below a salt concentration of about 1 M the depletion layer thickness was found to decrease with increasing polyelectrolyte concentration. At much lower salt concentrations the results of the SCF computations revealed a significant repulsive barrier in the interaction potential between two (uncharged) parallel flat plates. This is reminiscent of the findings of Mao \textit{et al}. \cite{mao1995depletion} who predicted such repulsive barriers for the interaction between two hard walls (or similarly for two hard spheres) in the presence of small hard spheres. 

The repulsive barrier in the effective pair interaction which exists for mixtures of large and small hard spheres, is further enhanced when both species are like-charged, as predicted by Walz and Sharma \cite{walz1994effect}. For low concentrations of the small charged spheres they predict that the depletion interaction is attractive for any salt concentration and that the strength of attraction at contact increases as the Debye length increases, or if the charge density on the large colloidal spheres increases. At higher concentrations of the small particles a repulsive barrier in the interaction potential curve appears for sufficiently large size ratios of small and large colloids and sufficiently large values of the Debye length. They could confirm many of the theoretical trends, including the repulsive barriers and oscillations in the interaction potential, by total internal reflection microscopy (TIRM) experiments on the force between a latex sphere and a glass wall in the presence of various nonadsorbing depletants. \cite{sober1995measurement,sharma1996direct,sharma1997measurement,sharma1997effect} 

These repulsive barriers have also been reported in mixtures of charged colloids and polyelectrolytes. Biggs \textit{et al}. \cite{biggs2000molecular} made direct atomic force microscopy measurements (AFM) of the force between two silica surfaces in sodium polystyrene sulfonate solutions. They reported a secondary minimum in the interaction potential, which became deeper on either increasing the polymer concentration or molar mass. Burns \textit{et al}. \cite{burns2002effect} found similar trends upon studying polystyrene latex particles under the influence of nonadsorbing polyacrylic acid using AFM. In case of neutral polymeric depletants the repulsive barriers are (nearly) absent \cite{Louis2002b}. They seem to appear only if the depletants are either hard particles or are charged, i.e., the self-interaction between the depletants must be sufficiently repulsive. 

Ferreira \textit{et al}.~\cite{ferreira2000mixtures} applied PRISM to study mixtures of charged polyelectrolytes and charged colloids, compared to a reference system of charged colloids and uncharged polymers. For charged polymers mixed with charged colloids PRISM predicts that the pair interaction between the colloidal spheres oscillates and is highly dependent on the parameters chosen. For high salt concentration (i.e., charges are screened) the potential at contact is repulsive due to double layer repulsions between the colloidal particles. The repulsion is found to be stronger if the polymers are neutral. At somewhat larger inter-particle distances there is an attractive part in the interaction potential, which is stronger for charged polymers compared to neutral polymers. In the presence of highly charged polymers the contact potential becomes attractive. At larger interparticle distances the potential then exhibits a repulsive barrier. \comment{DFT computations and simulations by J\"onssen \textit{et al}.~\cite{Jonsson2003} show that these structural oscillatory forces for charged spheres and polyelectrolytes are qualitatively not so sensitive to the magnitude and sign of the charge of the confining surfaces.}

Odijk \cite{odijk1997depletion} theoretically studied the depletion effects on the interaction between (small) colloidal particles interaction mediated by like-charged polyelectrolytes. He related the effective depletion thickness around the colloidal particles to the Debye length, the effective charge density at the surface of the colloidal spheres and to the Kuhn length. From this it appears that at high salt concentrations like-charges on polymers and colloids do not seem to affect the depletion induced attraction between colloids due to polyelectrolytes. At low ionic strength however the situation becomes quite complicated and detailed theories still have to be developed that enable a computation of the stability of such systems.  Also recent experimental and theoretical studies \cite{najafi2020synergistic,pryamitsyn2014interplay} show that the forces between charged colloids in the presence of polyelectrolyte solutions show nonmonotonic, sometimes oscillatory \cite{LUDWIG2020137,moazzami2016interplay} trends. 

In this article we show how a Donnan equilibrium \cite{Donnan1911} arises when polyelectrolytes are excluded from depletion zones. Unless high ionic strength screens the charges, this strongly affects the distribution of ions between the depletion zones and the rest of the system, and also provides a driving force for a \comment{spontaneous} shift of the polyelectrolyte towards the interplate region. \comment{We show that when a polyelectrolyte is allowed to deform at the cost of paying a conformational self-energy penalty, a competition between electrostatic and conformational penalties ensues, and a resulting equilibrium Donnan potential difference can be calculated from which we find disjoining pressures.} In certain limits we derive simplified analytic expressions that quantify the pressures. Our approach can be seen as the so-called zero-field limit \cite{Philipse2013}, valid when the Debye length is sufficiently large compared to the distances between the nonadsorbing surfaces and the electric field vanishes. Although this is arguably a rather simple model, it has previously yielded tractable predictions that could be verified experimentally \cite{Stojimirovic2020}.

The disjoining pressures predicted by our model are verified by comparison with numerical self-consistent field lattice computations in which electrostatic interactions are accounted for on a Poisson--Boltzmann level. Furthermore, we show under which conditions the match is expected to be quantitative, and in which cases a moderate deviation is found.



\section{Theory}

\subsection{Disjoining pressure at contact}\label{sec:deltapi-contact}
We start by calculating the disjoining pressure between two flat plates submerged in a polyelectrolyte solution, at close contact. In this case, the conformational self free energy of deformation of the polyelectrolytes is prohibitively high and the polyelectrolyte is entirely excluded from the zone between the flat plates, however, co- and counterions are small enough to enter the interplate space. We simplify the system by neglecting any gradients in the electric potential,  i.e. the  zero field or Donnan limit \cite{Donnan1911,Philipse2013}. \comment{In general, this assumption is valid when the plate separation is significantly smaller than the Debye length.} While imperfect, the Donnan limit often works over a larger range of conditions than expected, and has the additional advantage that the resulting models can be quite insightful. In this case, the concentrations of ions inside the interplate region are given by the Boltzmann distribution

\begin{equation}
    n_\pm^\text{i} = n_\pm^\text{res} \exp (\mp \Psi_\text{D}),\label{eq:internalConcentrations}
\end{equation}
where $n_\pm^\text{i}$ is the number density of positive and negative ions respectively between the plates, and $n_\pm^\text{res}$ the reservoir number density of positive and negative ions. Furthermore, $\Psi_\text{D} = e \psi_\text{D} / k_\text{B}T = \Psi^\text{i} - \Psi^\text{res}$ is the dimensionless Donnan potential in the interplate region. We assume both ideal ion behaviour, as well as a sufficiently large reservoir so that the reservoir concentration does not noticably change. As such, the reservoir concentrations of the ions are given by
\begin{subequations}
\begin{align}
    n_{+}^\text{res} &= n_\text{s}, \\
    n_{-}^\text{res} &= n_\text{s} + z n_\text{p},     
\end{align}
\end{subequations}
where $n_\text{s}$ is the number density of added monovalent salt, $n_\text{p}$ the reservoir number density of polyelectrolyte molecules, each carrying $z$ charges. In the remaining discussion we will tacitly assume that the polyelectrolyte carries positive charges --- when the charges on the polyelectrolyte are negative, the resulting equations remain valid, with only an exchange in the identities of the co- and counterions and a switch in sign of the resulting Donnan potential. The distribution of ions over the two regions is determined by the magnitude of the Donnan potential $\Psi_\text{D}$, which can be found by applying the electroneutrality condition, $n_{+}^\text{i} = n_{-}^\text{i}$, over the interplate region. The electroneutrality condition leads to the following charge balance equation
\begin{equation}
    n_\text{s} \exp (-\Psi_\text{D}) = ( n_\text{s} + z n_\text{p} ) \exp (\Psi_\text{D}),
\end{equation}
which, when solved for $\Psi_\text{D}$ results in
\begin{equation}
    \Psi_\text{D} = -\frac{1}{2} \ln \left(\frac{n_\text{s}+z n_\text{p}}{n_\text{s}}\right).\label{eq:closeContact}
\end{equation}

The osmotic pressure in the interplate region at close contact follows from the internal concentrations in \cref{eq:internalConcentrations} and the resulting Donnan potential in \cref{eq:closeContact}. Assuming Van 't Hoff's law, 
\begin{equation}
\frac{\Pi}{k_\text{B}T} = \sum_i n_i,
\label{eq:vantHoff}
\end{equation}
with $n_i$ the number density of species $i$, the osmotic pressure in the interplate region can be written as
\begin{equation}
    \frac{\Pi^\text{i}}{k_\text{B}T} = 2n_\text{s} \sqrt{1+\frac{zn_\text{p}}{n_\text{s}}}.
\end{equation}
The disjoining pressure equals the osmotic pressure difference of the interplate region with respect to the reservoir osmotic pressure, $\Delta \Pi = \Pi^\text{i} - \Pi^\text{res}$, where \comment{we assume that the reservoir osmotic pressure $\Pi^\text{res}$ due to both ions and polymer segments also follows from Van 't Hoff's law:}
\begin{align}
    \frac{\Pi^\text{res}}{k_\text{B}T} = (z+1) n_\text{p} + 2n_\text{s}.
\end{align}
\comment{For polymers, the Van 't Hoff osmotic pressure is an appropriate assumption in case of theta solvent conditions at low concentrations, which we adopt throughout for the sake of simplicity, but it should be noted that more complex models taking activity coefficients into account can and should be used in more general situations. This applies also to the ions, where deviations from ideality are to be expected at elevated concentrations.}

\subsection{Free energy of deformation in interplate region}
The Donnan potential between the plates bears a sign opposite to the sign of the polyelectrolyte valence, as follows from Eq.~\eqref{eq:closeContact}. As a consequence, the Donnan potential provides a driving force that shifts the polyelectrolyte density profile towards the interplate region, even when the interplate separation is smaller than the typical size of the polyelectrolyte chain. To account for this effect, we allow the polyelectrolyte to deform partially to enter the interplate space, at the cost of a free energy penalty incurred by the decreased conformational entropy associated with this deformation. This free energy penalty depends on the interplate separation, and is assumed here to be independent of the electrical potential between the plates. \comment{We expect this deformation to occur spontaneously as the system responds to the Donnan potential, provided the conformational free energy cost is not prohibitively high.} In this case, there is now a finite concentration of polyelectrolyte in the interplate region that contributes to the electroneutrality condition. This internal concentration $n_\text{p}^\text{i}$ is given by
\begin{equation}
    n_\text{p}^\text{i}(h) = n_\text{p} \exp \left\{ -[z \Psi_\text{D}(h) + \beta F_\text{p}(h)]\right\}.\label{eq:internalConcentration}
\end{equation}
Here, $\beta = (k_\text{B}T)^{-1}$ is the inverse thermal energy and $F_\text{p}(h)$ the (entropic) self-energy associated with a deformation to fit between two flat plates separated by a distance $h$. In the limit of $h\gg R_\text{g}$, when the interplate distance is far larger than the typical size of the polyelectrolytes, this conformational free energy penalty is zero and there is no concentration difference between the interplate and reservoir regions. In the limit of $h\rightarrow 0$ the free energy penalty diverges and the polyelectrolytes are fully excluded from the interplate region, as was assumed in \cref{sec:deltapi-contact}.

The canonical partition sum of a flexible polymer chain confined by two parallel flat plates was calculated by Casassa \cite{Casassa1967EquilibriumVoids}, the conformational part of which is given by
\begin{align}
  \chi(h) 
   &= 
   \frac{8}{\pi^2} \sum_{n = 1,3,5,\dots}^\infty \frac{1}{n^2}\exp \left(-n^2 \pi^2 \frac{R_\mathrm{g}^2}{h^2}\right), \label{eq:casassa}
\end{align}
\comment{where $R_\mathrm{g} = b\sqrt{M/6}$ is the polymer radius of gyration, with $b$ the polymer segment length and $M$ the number of segments per chain.} The conformational free energy penalty associated with this deformation is then calculated in a straightforward manner using $F_\text{p}(h) = - k_\text{B} T \ln \chi(h)$.

The resulting Donnan potential is a function of the interplate separation $h$, and can  be calculated from the condition of electroneutrality leading to the charge balance equation
\begin{equation}
    (n_\text{s}+z n_\text{p}) \exp \big[+ \Psi_\text{D}(h)\big] = n_\text{s} \exp \big[-\Psi_\text{D}(h)\big] + z n_\text{p} \chi (h) \exp \big[-z\Psi_\text{D}(h)\big].\label{eq:DonnanElectroNeutral}
\end{equation}
As mentioned before, this equation assumes that the charges on the polyelectrolyte are positive. When the charges on the polyelectrolyte are negative, only the sign of the Donnan potential switches.
For brevity, in the following we will drop the explicit dependencies of $F_\text{p}$ and $\Psi_\text{D}$ on $h$.

\begin{figure}[tbp]
\centering
\includegraphics[scale=\stw]{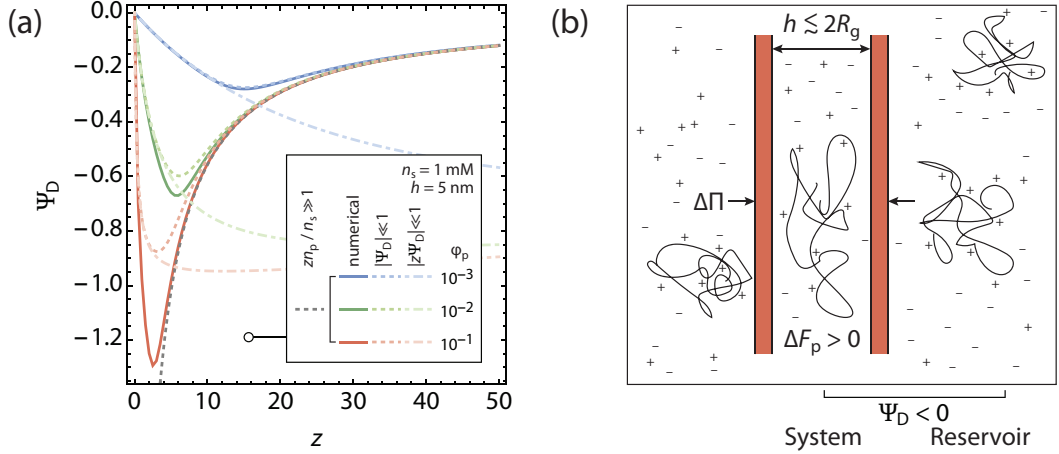}
\caption{{\bf{Donnan potential between flat plates.}} {\bf{a}} For a fixed interplate distance of $h=\SI{5}{\nano\meter}$ the self-energy for polyelectrolytes to enter the interplate region is $F_\text{p}(h)=\SI{6.1}{\kT}$. For higher polyelectrolyte valences, the self-energy is quickly overcome and the resulting Donnan potential collapses to a universal curve, independent of polyelectrolyte volume fraction. The solid curves denote the numerical solution to the charge balance \cref{eq:DonnanElectroNeutral}, whereas the dotted grey curve denotes the analytical approximation in the limit of $z n_\text{p}\gg n_\text{s}$ (\cref{eq:highZlimit}), the dotted coloured curves denote \cref{eq:lowPsi}, valid in the limit $|\Psi_\text{D}|\ll 1$, and the dotdashed lighter coloured curves denote \cref{eq:lowzPsi} which is valid in the limit $|z\Psi_\text{D}|\ll 1$. The number density of the polyelectrolyte is related to the volume fraction through $n_\text{p} = \phi_\text{p}/(M b^3)$ where $M=1000$ is the number of polymer segments and $b=\SI{0.3}{\nano\meter}$ is the segment length. \comment{{\bf{b}} Schematic of the double plate geometry, illustrating how deformation of polyelectrolyte allows a nonzero concentration in the depletion zones to yield lower overall free energy despite the entropic penalty of deformation.}}
\label{fig:DonnanPotential}
\end{figure}

In \cref{fig:DonnanPotential}a we show the Donnan potential, calculated numerically from \cref{eq:DonnanElectroNeutral} as a function of the polyelectrolyte valence $z$ at a number of different polyelectrolyte volume fractions at fixed inter-plate distance $h$, together with a schematic of the system under consideration in \cref{fig:DonnanPotential}b. At low $z$ the Donnan potential quickly becomes more negative but soon reaches a minimum. Surprisingly, the Donnan potential tends towards a single curve independent of polymer volume fraction for higher $z$. This can be seen by taking the limit of $z n_\text{p} \gg n_\text{s}$ of the charge balance \cref{eq:DonnanElectroNeutral}, which has the analytical solution 
\begin{equation}
    \Psi_\text{D} \simeq -\frac{\beta F_\text{p}}{1+z}.\qquad (z n_\text{p} \gg n_\text{s})\label{eq:highZlimit}
\end{equation}

In the limit of $|z \Psi_\mathrm{D}| \ll 1$, essentially the opposite limit as before, we can expand the exponentials in the charge balance \cref{eq:DonnanElectroNeutral} up to first order. In this limit we find the analytical solution
\begin{equation}
    \Psi_\text{D} \simeq -\frac{z n_\text{p}}{2 n_\text{s}} \frac{1-\exp(-\beta F_\text{p})}{1+\frac{z n_\text{p}}{2 n_\text{s}}+\frac{z^2 n_\text{p}}{2 n_\text{s}}\exp(-\beta F_\text{p})}.\qquad (|z \Psi_\mathrm{D}| \ll 1)\label{eq:lowzPsi}
\end{equation}

Another possible limit can be found by taking $|\Psi_\text{D}|\ll 1$ and expanding the two exponential terms independent of $z$ to first order, but leaving the third exponential term intact. An analytical solution can then be found that is accurate for both low $z$ and high $z$, provided the Donnan potential is low:
\begin{widetext}
\begin{equation}
    \Psi_\text{D}\simeq \frac{1}{z} W_0\left[  \frac{z^2 n_\text{p}}{2n_\text{s} + z n_\text{p}} \exp \left(\frac{z^2 n_\text{p}}{2n_\text{s} + z n_\text{p}}-\beta F_\text{p}\right) \right]-\frac{z n_\text{p}}{2n_\text{s}+z n_\text{p}},\qquad (|\Psi_\mathrm{D}| \ll 1)\label{eq:lowPsi}
\end{equation}
\end{widetext}
where $W_0$ is the principal branch of the Lambert $W$ function.
These limiting equations are useful in the appropriate conditions as shown in \cref{fig:DonnanPotential}, but since the effect of the Donnan potential leads to exponential changes in the concentrations, in the general case it will be more accurate to use the numerical result.

The osmotic pressures due to the ions and polyelectrolytes in the interplate and reservoir can now be calculated from the concentrations. In the reservoir, they follow directly from Van 't Hoff's law, \cref{eq:vantHoff}, however in the interplate region, the polyelectrolyte chains give rise to an additional entropic chain contribution. In \cref{sec:twoDerivations} we formally derive this contribution from the canonical partition sum for the interplate region, which leads to the expression
\begin{subequations}
\begin{align}
    \frac{\Pi^\text{i}}{k_\text{B}T} &= n_\text{p}^\text{i} \left(1+\frac{h}{\chi(h)}\frac{\partial \chi}{\partial h}\right) + n_+^\text{i} + n_-^\text{i} \nonumber \\
    \begin{split}
    &= n_\text{p} \exp (-z\Psi_\text{D}-\beta F_\text{p}) \left(1+z+\frac{h}{\chi(h)}\frac{\partial \chi(h)}{\partial h}\right) \\
    &\qquad{}+ 2 n_\text{s} \exp (-\Psi_\text{D}),
    \end{split}
    \label{eq:OsmoticPressureIn}\ \\
    \frac{\Pi^\text{res}}{k_\text{B}T} &= (z+1) n_\text{p} + 2n_\text{s}.\label{eq:OsmoticPressureOut}
\end{align}
\end{subequations}
The disjoining pressure is then simply the difference between the internal and reservoir osmotic pressures, $\Delta\Pi = \Pi^\text{i} - \Pi^\text{res}$. Note that, as before, in the equations presented here, we assume that the polyelectrolyte and ions show ideal behaviour. In \cref{sec:twoDerivations,sec:casassa} we present different derivations of the theory that allow one to include the effects of a finite ion size and a different interaction parameter.

\subsection{Self-consistent field lattice computations}
To evaluate the accuracy of the model derived in the previous section, we have performed Scheutjens--Fleer self-consistent field computations on a solution confined between two hard walls containing solvent, polyelectrolyte/polymer and salt ions. These computations take into account full Poisson--Boltzmann electrostatics (on a mean-field level) and assume a finite size for all components. In the following, we will focus on a brief description of the parameters employed for the computations, as self-consistent field theory already has been described in detail elsewhere\cite{fleer1993polymers,scheutjens1979statistical,bohmer1990weak,leermakers2005fundamentals}.

The self-consistent field computations were carried out using the Scheutjens--Fleer method, as implemented in the \texttt{sfbox} software package. The method considers polymer chains placed on a lattice on a mean-field level\comment{, thus fluctuations are not taken into account.} The computations were carried out on a lattice with a single gradient, terminated on either end with a hard wall. \comment{As such, these SCF computations can be thought of as being on the level of Flory--Huggins theory augmented with Poisson--Boltzmann electrostatics, with a gradient in concentration occurring only in the direction normal to the hard walls.} The number of lattice layers was varied and the disjoining pressure was computed by taking numerically the derivative of the resulting grand potential to the distance $h$ between the walls. 

A cubic lattice was used (corresponding with a lattice parameter of 1/6), where the size of each lattice site is $b = \SI{0.3}{\nm}$ with corresponding volume $b^3$. The polyelectrolyte chains were composed of $M = 1000$ segments, where each segment exactly occupies a single lattice site, thus its volume fraction is defined as $\phi_\mathrm{p} = n_\mathrm{p} b^3 M$. The solvent and ions all were composed of one segment.
The dielectric constant was set to $\epsilon = 80$ to model water.
The Flory--Huggins interaction parameter between polyelectrolyte and solvent was set to $\chi_\text{polymer--solvent} = 0.5$, between polyelectrolyte and wall to  $\chi_\text{polymer--wall} = +1$, whereas all other interaction parameters were set to zero ($\chi_\text{polymer--ions} = 0$, $\chi_\text{solvent--ions} = 0$,  $\chi_\text{solvent--wall} = 0$, and $\chi_\text{ions--wall} = 0$). The charge of the polyelectrolyte was smeared uniformly over all monomers unless otherwise stated.

\section{Results \& Discussion}

\begin{figure*}[tbp]
\centering
\includegraphics[scale=\stw]{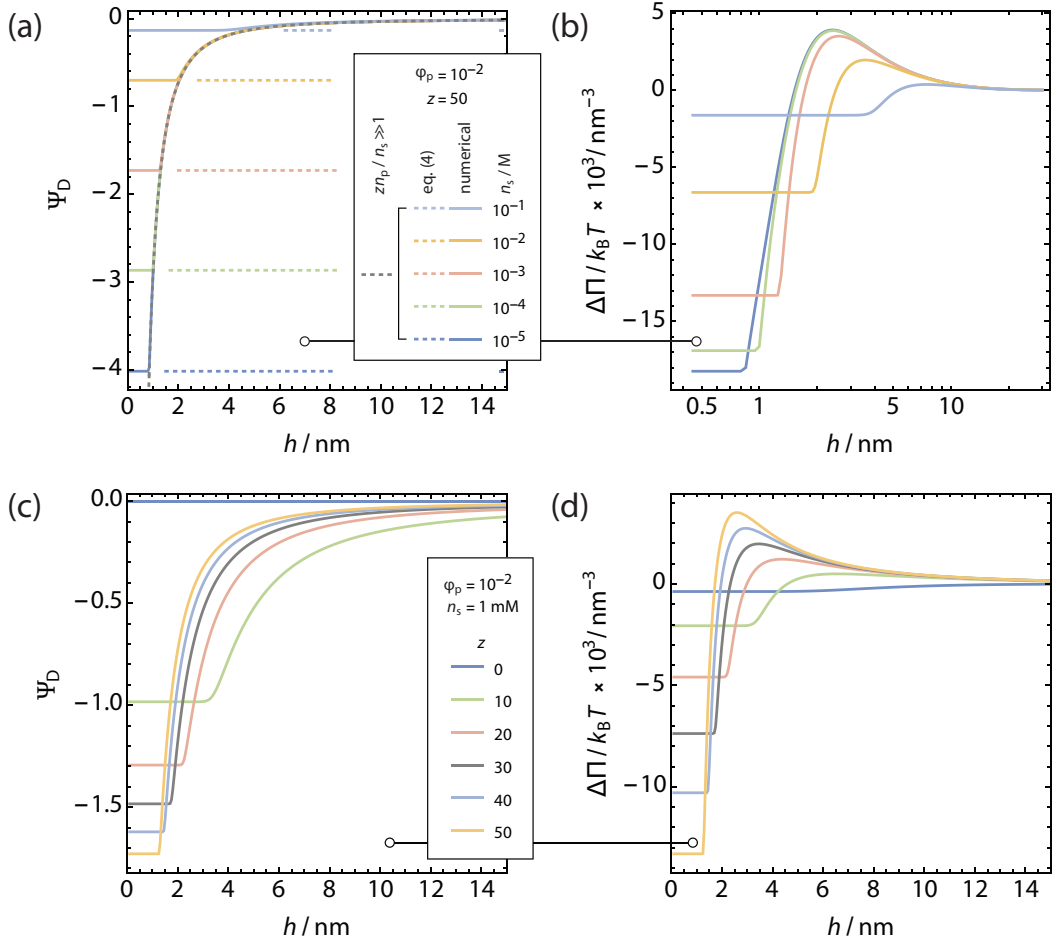}
\caption{{\bf{The effect of ionic strength and polyelectrolyte valence on the disjoining pressure.}} \lab{a} Donnan potential and \lab{b} disjoining pressure between flat plates as a function of interplate distance, for a number of different salt concentrations. \lab{c} Donnan potential and \lab{d} disjoining pressure between flat plates as a function of interplate distance, for a number of different polyelectrolyte valences. At small interplate distances, the self-energy for polyelectrolytes to enter the interplate region is prohibitively high, leaving a constant Donnan potential and disjoining pressure, given by \cref{eq:closeContact} (horizontal dashed line, \lab{a}). At intermediate distances the driving force for polyelectrolytes towards the interplate region is sufficiently high to overcome the self-energy, and a stabilising regime is found. Finally, at high interplate distances both the Donnan potential and the disjoining pressure tend towards their bulk values. The dotted grey curve in \lab{a} denotes the analytical approximation in \cref{eq:highZlimit}.}
\label{fig:DisjoiningPressure}
\end{figure*}

In \cref{fig:DisjoiningPressure} we plot the Donnan potential and disjoining pressure as a function of the inter-plate distance $h$ for a polyelectrolyte solution between two flat plates using \cref{eq:DonnanElectroNeutral,eq:OsmoticPressureIn,eq:OsmoticPressureOut} for varying ionic strength (\lab{a,b}) and varying polyelectrolyte valence (\lab{c,d}). From \cref{fig:DisjoiningPressure}\lab{a,c} we see that as the separation $h$ between the plates decreases, the Donnan potential becomes more negative, a consequence of the increasing conformational free energy penalty for the polyelectrolyte to enter the interplate region. The universal behavior, as predicted by \cref{eq:highZlimit}, is clearly visible in \lab{a}. When the driving force for polyelectrolyte chains to move towards this region is insufficient to overcome the entropic penalty, the Donnan potential becomes constant, regaining the result at close contact in \cref{eq:closeContact} (dashed horizontal lines in \cref{fig:DisjoiningPressure}\lab{a}). Here the presence of polyelectrolyte in the interplate region becomes negligible. The Donnan potential and interplate separation $h$ where this transition occurs depends on the polyelectrolyte valence and the salt concentration in the solution. In general, both lower ionic strengths and higher polyelectrolyte valences lead to stronger (more negative) Donnan potentials at close contact and a concurrent smaller interplate distance where this transition occurs.

\begin{figure*}[tbp]
\centering
\includegraphics[scale=\stw]{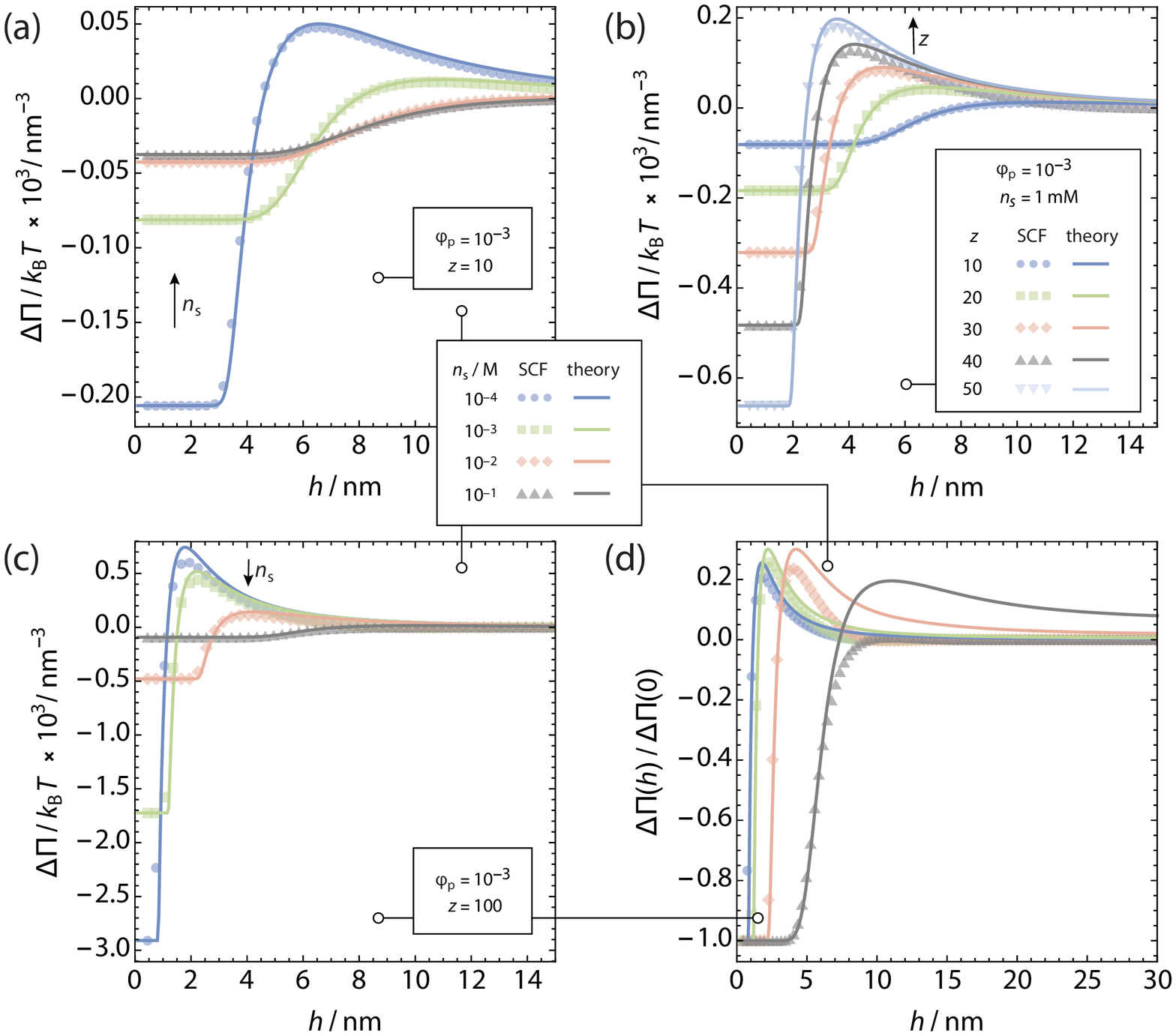}
\caption{{\bf{Comparison of the Donnan theory with numerical SCF calculations.}} Disjoining pressures are shown as a function of interplate separation $h$ for a number of salt concentrations (\lab{a, c, d}) and valences (\lab{b}). For low polyelectrolyte valence (\lab{a}, $z=10$) the Donnan theory provides an excellent quantitative match with SCF calculations of the same geometry, over a wide range of salt concentrations. \lab{b} As the polyelectrolyte valence increases, the Donnan theory starts to overestimate the stabilisation regime slightly, although the match is still quantitative. Moreover, the transition between depletion and enrichment of polyelectrolyte is described remarkably well. For a far higher polyelectrolyte valence (\lab{c}, $z=100$) the overestimation of the stabilisation regime becomes worse. 
When the disjoining pressure is normalised by the disjoining pressure at contact in \lab{d}, it becomes apparent that Donnan theory shows relatively the largest errors at high salt concentrations, where the effects electrostatic screening become increasingly important. 
}
\label{fig:CompSCF}
\end{figure*}

In \cref{fig:DisjoiningPressure}\lab{b,d} we show the disjoining pressure caused by the partitioning of (poly)ions corresponding to the Donnan potentials shown in panels \lab{a,c}. The disjoining pressure at close plate contact follows the trends of the Donnan potentials, with a transition to constant behaviour where the concentration of polyelectrolyte in the interplate region becomes negligible. Strikingly, we observe a maximum in the disjoining pressure due to the presence of polyelectrolyte between the plates --- at these interplate separations, the interaction between the plates is actually repulsive. The height of the maximum follows the same trends as the depth of the disjoining pressure at close contact, where lower ionic strength and higher polyelectrolyte valences lead to a higher maximum in the disjoining pressure. 

The repulsive forces between the plates are not a consequence of an electric double layer repulsion --- after all, the electric double layer is not taken into account in this zero-field model. Instead, the repulsion arises directly from the effect of confining the polyelectrolytes in a narrow space, and is due to the reduced conformational entropy concurrent with this confinement. While it would be entropically favourable, the enthalpic cost of a polyelectrolyte leaving the interplate space is prohibitively high, scaling with the exponent of $z \Psi_\text{D}$ as can be seen from \cref{eq:internalConcentration}. As a result, the polyelectrolyte concentration remains essentially constant until small separations, and the polyelectrolyte chains effectively act as entropic springs between the plates.

\comment{In the limit of $z$ approaching zero, the model still allows neutral polymers to enter the depletion region. However the conformational free energy cost is prohibitively high in the absence of a driving force provided by the Donnan potential. As such, the concentration of neutral polymer in the depletion regions becomes negligible and the classical behaviour of depletion interactions with neutral polymers is recovered.}


In \cref{fig:CompSCF} we show comparisons between Donnan theory and SCF calculations. In general, the theory is able to quantitatively predict the disjoining pressure at close contact of the plates, as well as the sharp transition described above. Furthermore, the maximum in the disjoining pressure is accurately represented by the theory, and depending on the conditions the match is quantitative. For relatively low polyelectrolyte valence ($z=10$) the match between SCF and Donnan theory is excellent. With increasing $z$ the theory starts to overestimate the maximum. As can be seen from \cref{fig:CompSCF}\lab{c} this overestimation is larger in absolute sense at low ionic strength. However, upon normalising the disjoining pressure with respect to its value at close contact, we see that the relative deviation is actually larger and extends over a further range in the case of high ionic strength. In this salt concentration regime, the plate separation $h$ exceeds the Debye length, and consequently the assumption that the Donnan limit is valid --- i.e., there are no gradients in the potential --- essentially breaks down.  However, since the absolute effect of charge on the disjoining pressure in this region is quite small, this does not detract from the value of the theoretical predictions.

\subsection{The range of the repulsive regime}

\begin{figure*}[tbp]
\centering
\includegraphics[scale=\stw]{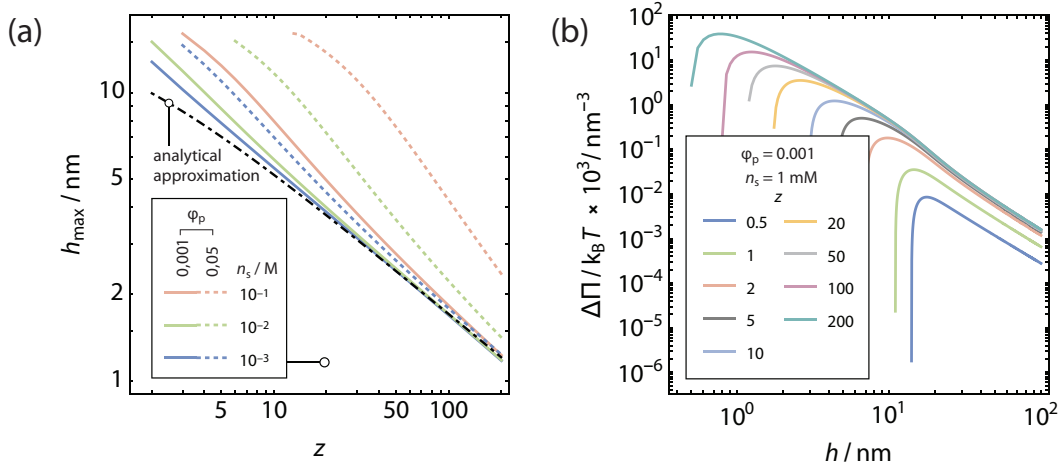}
\caption{{\bf{The range of the repulsive regime.}} \lab{a} The interplate separation at maximum repulsion $h_\text{max}$, plotted as a function of the polyelectrolyte valence, tends towards a universal curve as the valence grows. The effect is stronger when the salt concentration is low, or the polyelectrolyte concentration is high. The black dot-dashed line shows the analytical result in the limit $z n_\text{p}\gg n_\text{s}$ presented in \cref{eq:hMaxAnalytical}. \lab{b} The repulsive part of the disjoining pressure decays almost independently of the polyelectrolyte valence, except when the valence becomes vanishingly small. The main difference is that for higher polyelectrolyte valence, the transition to an interplate region depleted of polyelectrolyte occurs at narrower interplate separations.}
\label{fig:RepRange}
\end{figure*}

In \cref{fig:RepRange} we plot the interplate separation at maximum repulsion $h_\text{max}$ as a function of polyelectrolyte valence $z$ for a number of different ionic strengths and polyelectrolyte concentrations. Qualitatively, an increase in the polyelectrolyte valence causes the maximum in the repulsion to shift closer to contact --- a direct effect of the increased affinity for polyelectrolytes to remain between the plates even as the plates significantly confine any present polyelectrolyte. 

Surprisingly, as the valence increases, the curves tend to collapse to a universal behaviour, an effect that is more pronounced when the ionic strength is low or the polyelectrolyte volume fraction is high. In fact, this is the same limit $z n_\text{p} \gg n_\text{s}$ where we earlier observed a similar collapse in the resulting Donnan potential. To see this, we take the derivative of the disjoining pressure with respect to the interplate separation $h$ and set it to zero. Assuming that in this limit, counterions dominate the osmotic pressure behaviour over salt ions, we neglect the terms linear in $n_\text{s}$. As such, we get

\begin{widetext}
\begin{equation}
    \frac{\partial \Delta \Pi / k_\text{b}T}{\partial h} \simeq n_\text{p} \frac{\partial}{\partial h}\left[\exp (-z\Psi_\text{D}-\beta F_\text{p}) \left(1+z+\frac{h}{\chi(h)}\frac{\partial \chi(h)}{\partial h}\right) - (z-1)\right]=0.
\end{equation}
\end{widetext}

The final term, coming from the reservoir contribution, is independent of $h$ and does not survive the differentiation. Taking the limit $z n_\text{p} \gg n_\text{s}$ and substituting \cref{eq:highZlimit} for $\Psi_\text{D}$, we can evaluate the derivative, resulting in
\begin{widetext}
\begin{equation}
    n_\text{p}\exp(-z \Psi_\text{D})\left[\frac{\partial^2 \chi(h)}{\partial h^2}h+\frac{\partial \chi(h)}{\partial h}\left(2-\frac{z}{1+z}\frac{h}{\chi(h)}\frac{\partial \chi(h)}{\partial h}\right)\right] = 0.\label{eq:hderivative}
\end{equation}
\end{widetext}
Since the factor outside the brackets will be nonzero for physical values of $\Psi_\text{D}$, we can divide the factor out. The factor within brackets can be solved for $h$ numerically, however, we can proceed analytically by truncating Casassa's function \cref{eq:casassa} to only a single term. This is acceptable, because the maximum typically occurs for $h \lesssim R_\mathrm{g}$ for highly charged polyelectrolytes. In that case, \cref{eq:hderivative} has two analytical solutions, of which one is positive and physical:
\begin{align}
    h_\text{max} 
     &= R_\mathrm{g} \sqrt{\frac{2\pi^2}{1+z}}
    . \label{eq:hMaxAnalytical}
\end{align}
Even after truncating Casassa's function to a single term, the resulting solution is surprisingly accurate, and matches quantitatively with the numerical solution to \cref{eq:hderivative}, as we show in \cref{sec:truncatingCasassa}. In \cref{fig:RepRange} we plot the analytical solution in \cref{eq:hMaxAnalytical} (black dashdotted line). It is clear that the interplate separation at maximum repulsion $h_\text{max}$ tends towards the analytical solution in the limit where counterions dominate the behaviour in the interplate region. \Cref{eq:hMaxAnalytical} can also be interpreted as a typical electrical length scale, related to the radius of gyration of the polyelectrolyte, modified by its valence.

\subsection{Distribution of charge along polyelectrolyte}


\begin{figure}[tbp]
\centering
\includegraphics[scale=\stw]{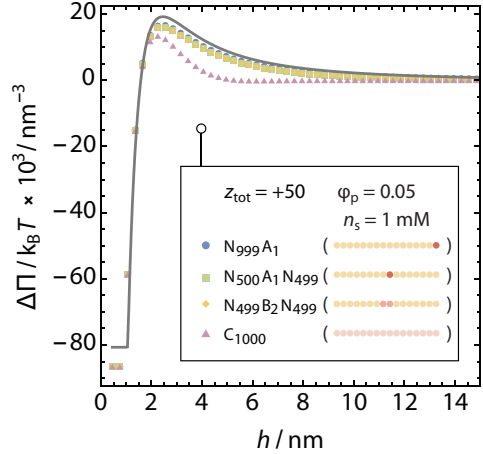}
\caption{{\bf{Effect of the charge distribution on the polyelectrolyte.}} Disjoining pressure as a function of interplate separation. The Donnan theory is compared to SCF calculations where the charge was placed on a single segment at the end of a polymer (\ce{N999A1}), on one or two segments near the middle of the polymer (\ce{N500A1N499} and \ce{N499B2N499}) and smeared out over all segments (\ce{C1000}). Here, \ce{N} indicates a neutral segment, \ce{A} a segment bearing a charge of \num{+50}, \ce{B} bearing a charge of \num{+25} and \ce{C} bearing a charge of \num{0.05}. Consequently, the total charge on the polymer was kept constant at \num{50} for all calculations. The Donnan theory describes more accurately the SCF calculations where charge is concentrated on a small part of the polymer.}
\label{fig:ChargeDistribution}
\end{figure}

One explanation for the overestimation of the Donnan theory with respect to numerical SCF calculations, apart from neglecting effects of electrostatic screening,  is that the Donnan theory essentially assumes the polyelectrolyte to be a point charge superimposed with a neutral polymer chain. We investigated the effect of this assumption using SCF calculations. In \cref{fig:ChargeDistribution} we compare the Donnan theory with SCF calculations where the total polyelectrolyte valence was distributed over the segments in four different ways: first, we concentrated all the charges on a single segment located at one end of the chain (\ce{N999A1}). Second, we concentrated all charge on a single segment in the middle of the chain (\ce{N500A1N499}), with a third variant where the total charge was smeared over the two middle segments (\ce{N499B2N499}). Finally, we compared this to the reference situation where the total charge is smeared out over the whole polymer (\ce{C1000}). We see that when the charge is concentrated locally, the maximum in disjoining pressure increases when compared to the situation where the charge is smeared out. For this effect there is a negligible influence whether this single segment is located at the end of a polymer chain or in the middle. The Donnan theory still somewhat overestimates the magnitude of the maximum, yet the difference is minute in this case.

\begin{figure*}[tbp]
\centering
\includegraphics[scale=\stw]{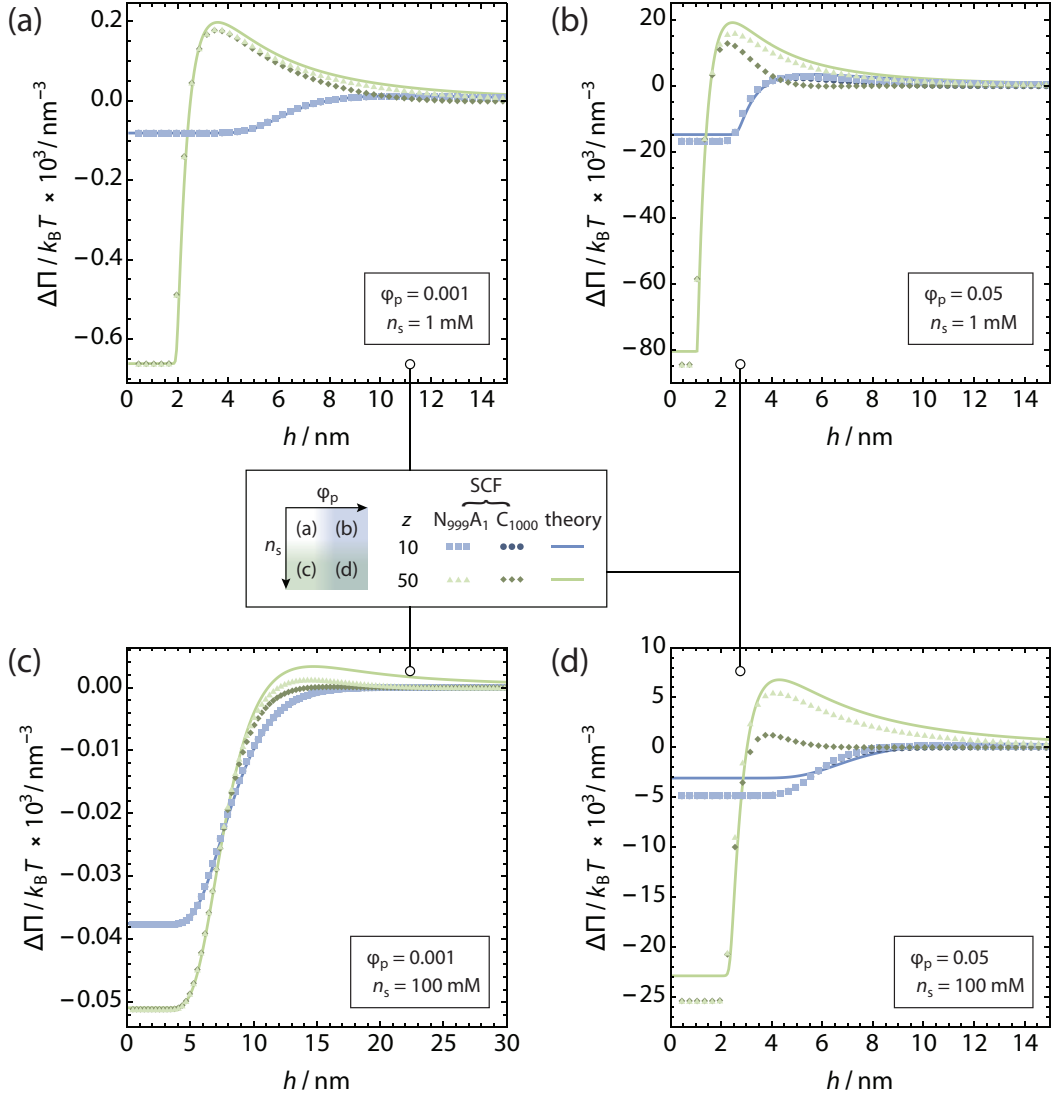}
\caption{{\bf{Regimes where the Donnan theory is applicable.}} Disjoining pressures as a function of the interplate separation $h$ for two different polyelectrolyte valences $z$ for low $\phi_\text{p}$, low $n_\text{s}$ (\lab{a}), high $\phi_\text{p}$, low $n_\text{s}$ (\lab{b}), low $\phi_\text{p}$, high $n_\text{s}$ (\lab{c}) and high $\phi_\text{p}$, high $n_\text{s}$ (\lab{d}). The Donnan theory is compared to SCF calculations where the charge is concentrated on a single segment (light symbols) or smeared out over the whole polymer (dark symbols). In all cases, the Donnan theory describes more accurately the SCF calculations where charge is concentrated on a single segment, although the differences are only visible for relatively high ionic strength (including counterions). Furthermore, when the polyelectrolyte volume fraction is very high, the disjoining pressure at contact also deviates.}
\label{fig:Regimes}
\end{figure*}

In \cref{fig:Regimes} we show in which conditions quantitative agreement between Donnan theory and SCF calculations can be expected, and whether significant difference can be expected between the behaviour of a polyelectrolyte with its charge concentrated on a single segment (\ce{N999A1}) and the behaviour of a polyelectrolyte with its charge fully smeared out (\ce{C1000}). In general, at low polymer and salt concentrations (panel \lab{a}), agreement between Donnan theory and SCF calculations is excellent. At higher polymer volume fractions (panel \lab{b}) there is noticable difference between the \ce{N999A1} and \ce{C1000} cases. A higher salt concentration (panel \lab{c}), in contrast, causes the Donnan theory to overestimate the maximum in disjoining pressure, and both effects are present when the concentrations of both polymer and salt are high (panel \lab{d}). In the latter case, quantitative agreement between Donnan theory and SCF calculations is moderate. It should be noted here that the concentration of counterions is by far the dominant contribution to the ionic strength in panels \lab{b,d}, where the polyelectrolyte volume fraction is high. One consequence of this is that the \num{100}-fold increase in salt concentration upon going from panel \lab{b} to \lab{d} has only a minor effect on the magnitude of the disjoining pressure. In contrast, upon going from panel \lab{a} to panel \lab{c}, the ionic strength is significantly affected by the addition of salt in the reservoir, and as such the disjoining pressure is decreased by more than a factor \num{10}. 

\section{Conclusions}

In mixtures of colloids and polyelectrolytes, a Donnan potential arises across the region depleted of polyelectrolyte. This Donnan potential provides a driving force for the polyelectrolytes to shrink the depletion region, and accumulate locally. In this paper, we derive a theoretical model that enables us to calculate the depletion interaction between two flat plates that arises due to polyelectrolyte in the reservoir. We allow the polyelectrolyte to enter the depletion region, at the cost of a conformational free energy based on Casassa's expression for the conformational partition sum of a Gaussian chain confined between flat plates \cite{Casassa1967EquilibriumVoids}. In the so-called zero-field limit, valid when the Debye length is sufficiently large compared to the distances between the non-adsorbing surfaces, the electric field vanishes. In such a case the partitioning of (poly)ions is governed by the requirement of electroneutrality across the system. The electroneutrality condition leads to a charge balance that can be solved numerically for the resulting Donnan potential. We also derive three analytical expressions for the Donnan potential in different limits.

The Donnan potential enables to calculate the disjoining pressure between the flat plates due to the presence of polyelectrolyte. We find that the disjoining pressure is constant at small interplate separations, and follows from the analytical limit when polyelectrolytes are forbidden from entering the interplate region. This disjoining pressure mediated by polyelectrolytes is significantly larger in magnitude as for neutral polymers. 

When the separation between the plates becomes comparable to the size of the polyelectrolyte chains, the disjoining pressure transitions suddenly to a repulsive regime. In this regime, polyelectrolytes confined between the plates exert an outward pressure due to the compressed entropic spring effect of the chains. The repulsive regime becomes larger with increasing polyelectrolyte valence and decreasing reservoir ionic strength --- either effect provides a stronger driving force for polyelectrolytes to enter the interplate region at smaller separations.

For low polyelectrolyte and ion concentrations, we find a quantitative match between our model and numerical self-consistent field lattice computations on a polyelectrolyte solution confined between two hard walls, which take into account full Poisson-Boltzmann electrostatics. For higher concentrations, the model still quantitatively predicts the crossover from the regime of close plate contact to the repulsive regime. We see that at high polyelectrolyte valence, when counterions contribute significantly to the ionic strength, the analytic Donnan theory overestimates the repulsive regime in comparison to the self-consistent field computations. A similar increase in the repulsive regime is found for self-consistent field computations where the charge on the polymer chains is concentrated on a single segment, rather than smeared out over the full chain.

\begin{acknowledgments}
MV acknowledges the Netherlands Organisation for Scientfic Research (NWO) for a Veni grant (no.~722.017.005) and Nerd and Et for useful discussions. The authors acknowledge prof.~Frans Leermakers for the \texttt{sfbox} software package.

\end{acknowledgments}

\section*{Data availability}
The data that support the findings of this study are available from the corresponding author upon reasonable request.

\appendix

\section{Concentration in the interplate region}

\begin{figure}[tbp]
\centering
\includegraphics[scale=\stw]{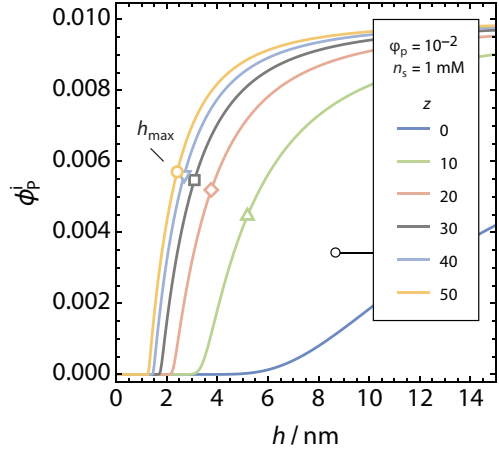}
\caption{{\bf{In spite of the repulsive maximum in the disjoinin pressure between the plates, the polyelectrolytes are always partially depleted from the interplate region.}} By plotting the concentration of polyelectrolytes in the interplate region as a function of interplate separation, it is apparent that the concentration always monotonically drops from its bulk value to $0$ at close contact. The repulsive barrier around $h = h_\mathrm{max}$ arises purely from the entropic spring confinement of the polyelectrolytes. This can also be observed from the position of the maximum in the disjoining pressure, indicated with an open symbol, which tends to occur when the interal polyelectrolyte volume fraction is about half of the bulk volume fraction. This maximum is calculated using \cref{eq:hMaxAnalytical}.}
\label{fig:segmentDensity}
\end{figure}

In the repulsive regime, quite a significant fraction of polyelectrolyte is already depleted from the interplate region. To see this, we show in \cref{fig:segmentDensity} the polyelectrolyte segment volume fraction between the plates, as a function of interplate separation. The figure also shows the maximum in disjoining pressure (symbol) for each curve, as calculated using \cref{eq:hMaxAnalytical}. The polyelectrolyte segment volume fraction between the plates is always equal or lower than the bulk volume fraction.

\section{Two extended derivations in the canonical and grand canonical ensemble}\label{sec:twoDerivations}

In this section, we present two more extended derivations of our model. The first, in the canonical ensemble, provides more detail to the shorter derivation in the main text. The second is an alternative in the grand canonical ensemble, which takes into account the finite size of the ions. As we will see, both provide near-identical results.

Starting in the canonical ensemble, we write down the canonical partition sum of the system between two flat plates at separation $h$,
\begin{equation}
    Z(h)  = \frac{z_\text{p}^{N_\text{p}^\text{i}}}{N_\text{p}^\text{i}!} \frac{z_+^{N^\text{i}_+}}{N^\text{i}_+!} \frac{z_-^{N^\text{i}_-}}{N^\text{i}_-!},
\end{equation}
where $N^\text{i}_\text{p},N^\text{i}_+,N^\text{i}_-$ are the numbers of particles of polyelectrolyte, positive and negative ions respectively inside the volume enclosed by the two plates, and $z_\text{p},z_+,z_-$ the individual particle canonical partition sums, which in this case are given by
\begin{subequations}
\begin{align}
   z_\text{p} &= V \exp (- z\Psi_\text{D}) \chi (h), \\
   z_+ &= V \exp (-\Psi_\text{D}), \\
   z_- &= V \exp (+\Psi_\text{D}).
\end{align}
\end{subequations}
Here, $V$ is the volume enclosed by the two plates, $z$ the polyelectrolyte valence, $\Psi_\text{D}$ the dimensionless Donnan potential that arises from the partial exclusion of polyelectrolyte from the region enclosed by the plates, and $\chi(h)$ is Casassa's conformational part of the canonical partition sum of a Gaussian chain confined between two flat plates at a distance $h$, given by
\begin{equation}
  \chi(h) = \frac{8}{\pi^2} \sum_{n = 1,3,5,\dots}^\infty \frac{1}{n^2}\exp \left(-\frac{n^2 \pi^2 b^2 M}{6 h^2}\right),\label{eq:casassa2}
\end{equation}
where $b$ is the polymer segment length and $M$ the number of segments per chain. We use Stirling's approximation to calculate the Helmholtz free energy from the partition sum, resulting in
\begin{align}
    \frac{F}{k_\text{B}T} &= N_\text{p}^\text{i}\left[\ln \frac{N_\text{p}^\text{i}}{V} - 1 - \ln \chi(h) + z \Psi_\text{D}\right] \nonumber \\
    &\quad + N^\text{i}_+ \left[\ln \frac{N^\text{i}_+}{V}-1+\Psi_\text{D}\right] \nonumber \\ 
    &\quad + N^\text{i}_- \left[\ln \frac{N^\text{i}_-}{V}-1-\Psi_\text{D}\right],\label{eq:FreeEnergy}
\end{align}
from which we can calculate the osmotic pressure between the plates as 
\begin{align}
    \frac{\Pi}{k_\text{B}T} &= -\frac{\partial (F/k_\text{B}T)}{\partial V} \nonumber \\
    &= n^\text{i}_\text{p} \left[1+\frac{h}{\chi(h)}\frac{\partial \chi(h)}{\partial h} - hz\frac{ \partial \Psi_\text{D}}{\partial h}\right]\nonumber \\ 
    &\quad +n^\text{i}_+ \left[1-h\frac{\partial \Psi_\text{D}}{\partial h}\right] +n^\text{i}_- \left[1+h\frac{\partial \Psi_\text{D}}{\partial h}\right].
\end{align}
Here we have converted the particle numbers to number densities through $n = N/V$. By requiring the system to be electrically neutral, we see that the terms dependent on $(\partial \Psi_\text{D} / \partial h)$ all cancel out, as $n^\text{i}_- = n^\text{i}_+ + z n^\text{i}$. As such we are left with the simpler form
\begin{equation}
    \frac{\Pi}{k_\text{B}T} = n^\text{i}_\text{p} \left[1+\frac{h}{\chi(h)}\frac{\partial \chi(h)}{\partial h}\right] + n^\text{i}_+ + n^\text{i}_-.
\end{equation}

The internal concentrations $n_\text{p}^\text{i}, n^\text{i}_+,n^\text{i}_-$ can be found from the chemical potentials that are calculated in the usual way from \cref{eq:FreeEnergy}. By imposing chemical equilibrium with the reservoir surrounding the two plates, the internal concentrations can be expressed in terms of the reservoir concentrations and the resulting Donnan potential.
\begin{subequations}
\begin{align}
    n_\text{p}^\text{i} &= n_\text{p} \exp \big[-z \Psi_\text{D} \chi(h)\big], \\
    n^\text{i}_+ &= n^\text{res}_+ \exp \big(- \Psi_\text{D}\big) = n_\text{s} \exp \big(- \Psi_\text{D}\big), \\ 
    n^\text{i}_- &= n^\text{res}_- \exp \big(+ \Psi_\text{D}\big) = (n_\text{s}+ z n_\text{p}) \exp \big(+ \Psi_\text{D}\big),
\end{align}
\end{subequations}

The Donnan potential in turn can be found by imposing charge balance on the interplate-system and the reservoir. The resulting charge balance equation, 
\begin{equation}
    (n_\text{s}+z n_\text{p}) \exp (+ \Psi_\text{D}) = n_\text{s} \exp (-\Psi_\text{D}) + z n_\text{p} \chi (h) \exp (-z\Psi_\text{D}),\label{eq:DonnanElectroNeutral2}
\end{equation}
needs to be solved numerically to obtain the appropriate $\Psi_\text{D}$.

An alternative derivation can be devised, making use of the grand canonical ensemble. The derivation is essentially lattice based: the interspace volume is divided into lattice sites with a cell volume $v = b^3$ set by the course grained length scale in the system. In this case it makes sense to set $b$ equal to the polymer segment length. The volume of the cell is chosen such that a cell can only be occupied by a single polymer segment, ion or solvent molecule, but all species can freely move between cells. In that case, a single lattice site can be seen as a grand canonical ensemble in its own right, with a grand canonical partition function given by
\begin{align}
    \Xi_\text{site} &= \sum_{N_\text{p}} \sum_{N_+} \sum_{N_-} \lambda_\text{p}^{N_\text{p}} \lambda_{+}^{N_+} \lambda_{-}^{N_-} Z(N_\text{p},N_+,N_-).
\end{align}

Here, $N_\text{p},N_+,N_-$ are the single-site copy numbers of polymers, co-ions and counterions respectively, which by virtue of the lattice-based setup can either be \num{0} or \num{1}. Furthermore, each species has a fugacity $\lambda_i = \exp (\beta \mu_i)$, with $\beta = (k_\text{B}T)^{-1}$ the inverse thermal energy. $Z(N_\text{p},N_+,N_-)$ is the (configurational part of the) canonical partition sum of the site, and since we only allow a single species to occupy the lattice site, this canonical partition sum can be expressed as the Boltzmann factor of the excess free energy $\epsilon_i$ associated with species $i$ occupying the lattice site. In case of the co- and counterions, we assume zero field and neglect any interactions between ions. As such, their excess free energy is entirely due to their interaction with the current Donnan potential $\Psi_\text{D}$. 

We have chosen to model the polyelectrolyte as a point charge with valence $z$ occupying the site, surrounded by a Gaussian chain. As such, the model takes into account the finite volume occupied by the ions, but underestimates the volume occupied by polymer segments and polymer-polymer excluded volume interactions are not taken care of well in this model. We do take into account the effect of the confinement by the two flat plates surrounding the system volume by including Casassa's conformational part of the canonical partition sum shown in \cref{eq:casassa2}. We can then write for all allowed configurations
\begin{equation}
\left\{\begin{aligned}
    Z(0,0,0) &= 1, \\
    Z(1,0,0) &= \exp (- z \Psi_\text{D}) \chi(h), \\
    Z(0,1,0) &= \exp (- \Psi_\text{D}), \\
    Z(0,0,1) &= \exp (+\Psi_\text{D}).
\end{aligned}\right.
\end{equation}

The chemical potentials of the species are set by the (much larger) reservoir. Assuming ideal solution behaviour, $\lambda_+ = \phi_\text{s}$, $\lambda_- = \phi_\text{s} + \phi_\text{p}z/M$ and $\lambda_\text{p} = \phi_\text{p}/M$, where $\phi_\text{s}$ is the volume fraction of salt in the reservoir and $\phi_\text{p}$ the volume fraction of polyelectrolyte segments in the reservoir. Finally, for the entire interplate-system we assume that all sites are independent, so that the grand partition function of the system reads

\begin{equation}
    \Xi_\text{sys} = \Xi_\text{site}^{V/v}.
\end{equation}

We can calculate the relevant thermodynamic variables of the system by taking the appropriate derivatives of the grand potential $\Omega=-k_\text{B}T \ln \Xi_\text{sys}$. The Donnan potential can be found by imposing the electroneutrality condition, which leads to the same charge balance \cref{eq:DonnanElectroNeutral2}.

\begin{figure}[tbp]
\centering
\includegraphics[scale=\stw]{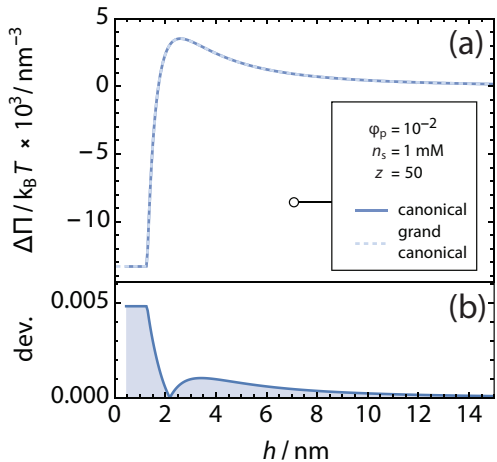}
\caption{{\bf{The difference between the canonical and grand canonical ensemble.}} \lab{a} Disjoining pressure as a function of interplate separation for the canonical and grand canonical ensemble for a representative set of conditions. As can be seen from the absolute deviation between the two results in \lab{b}, the two derivations can be used interchangably, yielding quantitative results with only very minor deviations.}
\label{fig:ensembleComparison}
\end{figure}

\section{Modified Casassa function}\label{sec:casassa}

Tuinier and Fleer \cite{tuinier2004concentration,tuinier2004bconcentration} derived a modification to Casassa's expression for the partition sum of a Gaussian chain confined between flat plates. The modification takes into account the changes in radius of gyration due to the Flory--Huggins interaction parameter between the polyelectrolyte chain and the solvent. Substituting the modified expression for the original in our model, we can calculate the effect of a different interaction parameter on the disjoining pressure (although it should be noted that we still assume the Van 't Hoff equation for the polymer osmotic pressure).

\begin{figure}[tbp]
\centering
\includegraphics[scale=\stw]{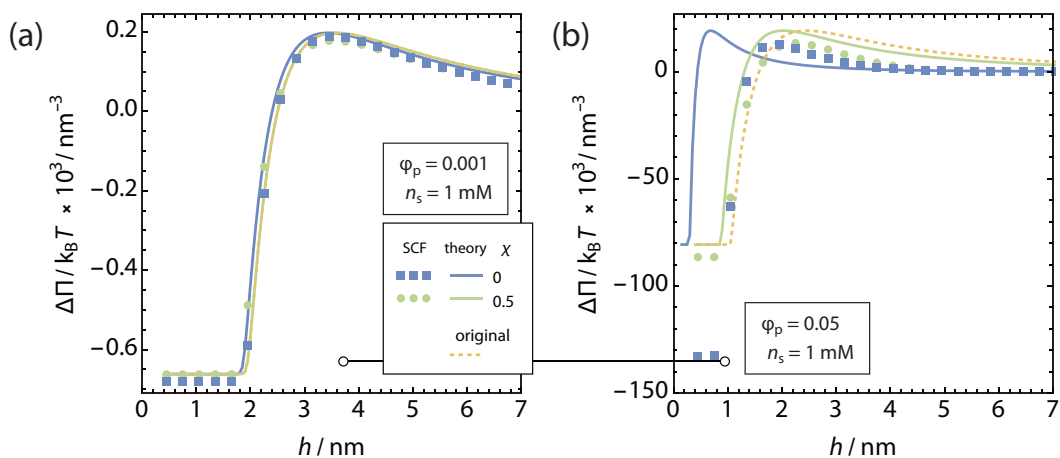}
\caption{{\bf{The modified Casassa function shifts the range of disjoining pressure curves depending on polymer solvency,}} \lab{a} for $\phi_\text{p} = 10^{-3}$ and \lab{b} for $\phi_\text{p} = 0.05$. Taking into account a different Flory--Huggins interaction parameter between solvent and polyelectrolyte can shift the range of the interaction between the plates. Especially at lower concentrations, for a solvent with $\chi=0.5$ the disjoining pressure is virtually identical to that obtained using the original Casassa function. However, for decreasing interaction parameter $\chi$ the curves shift to closer interplate separations. When the polyelectrolyte concentration is higher, a small shift can be observed, although the effect does not qualitatively change the behaviour. SCF calculations calculated with a solvent-polyelectrolyte Flory-Huggins parameter of $\chi = 0$ and $\chi = 0.5$ are provided for comparison, although at higher polyelectrolyte volume fractions, the quantitative match between SCF and the theory for $\chi = 0$ is lost.}
\label{fig:modifiedCasassa}
\end{figure}

In \cref{fig:modifiedCasassa} we show how changes to the Flory--Huggins interaction parameter $\chi$ between the polyelectrolyte and the solvent will shift the disjoining pressure curve. When the interaction parameter is reduced from $\chi=0.5$ to $\chi=0$, the disjoining pressure curve is compressed to smaller values of the interplate separation $h$. Essentially, the changed radius of gyration rescales the typical length scale in the system. The effect is far larger when the concentration of polyelectrolyte is higher, with barely any effect for $\phi_\text{p}=10^{-3}$, but a significant rescaling for $\phi_\text{p}=0.05$. This rescaling is far smaller in our SCF computations, which are shown as the data points in \cref{fig:modifiedCasassa}.

\section{Range of the repulsive effect}\label{sec:truncatingCasassa}

\begin{figure}[tbp]
\centering
\includegraphics[scale=\stw]{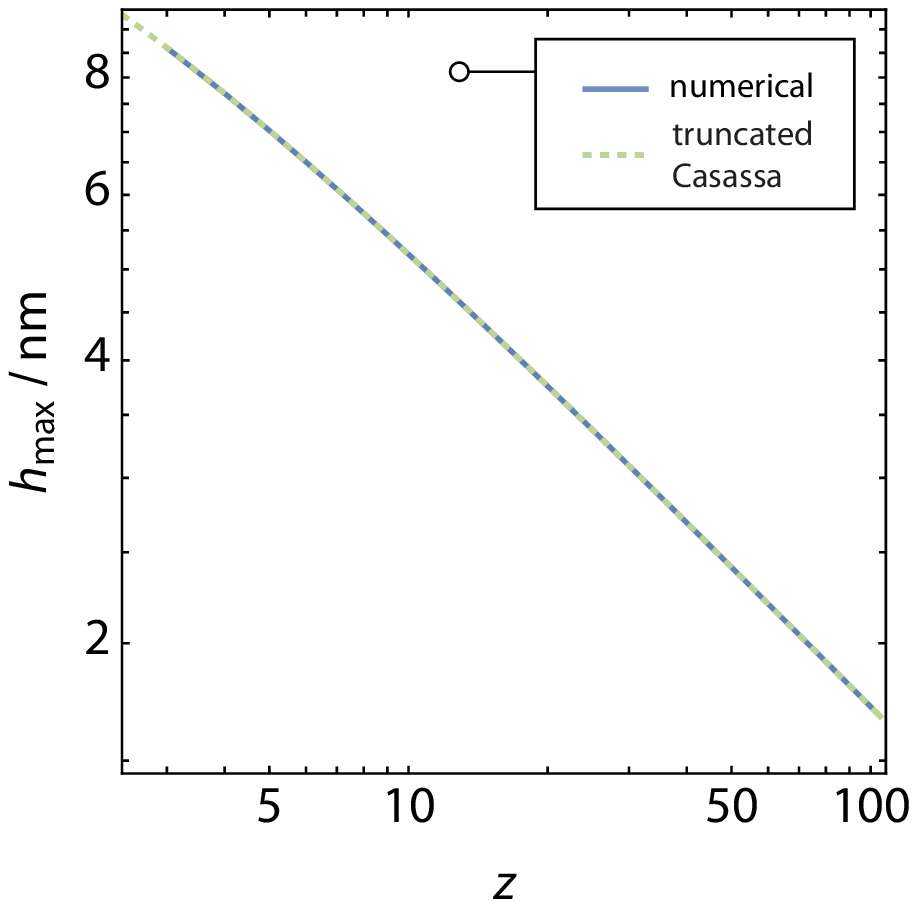}
\caption{{\bf{Truncating Casassa's equation provides a good analytical approximation.}} The maximum in the disjoining pressure can be calculated by numerically solving for the interplate separation $h$ where the derivative of the disjoining pressure is zero. Truncating Casassa's function to only a single term allows one to find the analytical expression of \cref{eq:hMaxAnalytical}, that matches quantitatively with the numerical solution to \cref{eq:hderivative}. Here we work in the limit of $z n_\text{p} \gg n_\text{s}$, where the curve becomes independent of salt concentration and polyelectrolyte volume fraction.}
\label{fig:SI-hmax}
\end{figure}

In \cref{fig:SI-hmax} we show that the analytical approximation of the maximum in disjoining pressure in \cref{eq:hMaxAnalytical}, obtained by truncating Casassa's function after its first term matches quantitatively with the numerical solution to \cref{eq:hderivative}. The truncation is acceptable since in the region where the maximum typically occurs, $h \lesssim R_\mathrm{g}$, the first term of Casassa's function dominates.

\bibliography{
refs_MV,refs_RT}

\end{document}